\documentclass{appolb}
\usepackage{
	epsfig
	,amsmath
	,graphicx
}

\begin{document}
\title{Diquark Bose-Einstein condensation at strong coupling
\thanks{Poster presented at the XXVI. Max Born Symposium
"Three Days of Strong Interactions", Wroc{\l}aw (Poland), 9.-11. VII. 2009}%
}
\author{
	D.~S.~Zab{\l}ocki
\address{
	Institute for Theoretical Physics, 
	University of Wroc{\l}aw, 50-204 Wroc{\l}aw, Poland\\
	Institute of Physics, University of Rostock, D-18051 Rostock, Germany\\
	Laboratory of Information Technologies, JINR Dubna, 
	141980 Dubna, Russia
}
	\vskip 1.em
	D.~B.~Blaschke
\address{
	Institute for Theoretical Physics,
	University of Wroc{\l}aw, 50-204 Wroc{\l}aw, Poland\\
	Bogoliubov Laboratory of Theoretical Physics, JINR Dubna, 
	141980 Dubna, Russia
}
	\vskip 1.em
	R.~Anglani
\address{
	Dipartimento di Fisica, Universit\`a di Bari, I-70126 Bari, Italia\\
	Istituto Nazionale di Fisica Nucleare, I-70126, Bari, Italia
}
\and\vskip -.5em
	Yu.~L.~Kalinovsky
\address{
	Laboratory of Information Technologies, JINR Dubna, 
	141980 Dubna, Russia
}
}
\maketitle
\begin{abstract}
We investigate the phase structure  of the
$SU_f(2)\otimes SU_c(3)$ Nambu--Jona-Lasinio model
as a function of the scalar diquark coupling strength.
Above a critical coupling, the binding energy is sufficiently large
to overcompensate the quark masses and a massless scalar diquark
bound state emerges which leads to Bose condensation already in the
vacuum.
\\[-17cm]
\flushright{
BA-TH/616-09
}
\\[17cm] 
\end{abstract}

\section{Introduction}

Recent laboratory experiments with ultracold gases of fermionic atoms
allow to investigate dense Fermi systems with their coupling strength
tunable via Feshbach resonances by applying external magnetic fields.
After the preparation of fermionic dimers in 2003,
now also their Bose-Einstein condensation (BEC)
\cite{Greiner:2003,Zwierlein:2003zz} and superfluidity of these dimers has
been observed \cite{Zwierlein:2005,Greiner:2003a}.
Besides this strong coupling regime, for weak attractive interactions
at low enough temperatures the condensation of bosonic correlations
(Cooper pairs) in the continuum of unbound states occurs
according to the Bardeen-Cooper-Schrieffer (BCS) theory.

The BEC-BCS crossover is physically related \cite{Calzetta:2006}
to the bound state dissociation or Mott-Anderson delocalization transition
\cite{Mott:1968zz} where the modification of the effective coupling strength is
caused by electronic screening and/or Pauli blocking effects.
The Mott transition is a very general effect expected to occur in a wide 
variety of dense Fermi systems with bound states
such as
deuterons in nuclear matter \cite{Schmidt:1990,Stein:1995,Schnell:1999tu}
or diquarks in quark matter
\cite{Kitazawa:2001ft,Blaschke:2004cs,Blaschke:2005uj,Ebert:1992ag}.
Below a critical temperature, bosonic correlations form a condensate and this
transition appears as BEC-BCS crossover, which in quark matter is of particular
theoretical interest due to the ultrarelativistic regime for massless 
(Goldstone) bosons \cite{Abuki:2006dv,Deng:2006ed,Sun:2007fc}.

A systematic treatment of these effects is possible within the path integral
formulation for finite-temperature quantum field theories.
This approach is rather general as it is relativistic and is especially suited
to take into account the effects of spontaneous symmetry breaking.
In this contribution we sketch the basics of this approach on the
example of a model field theory of the Nambu--Jona-Lasinio type for a
relativistic strongly interacting Fermi system, see \cite{Buballa:2003qv}
for a recent review.
These investigations are also motivated by the analogies of the strongly
coupled quark-gluon plasma (sQGP) at Relativistic Heavy Ion Collider (RHIC)
in Brookhaven \cite{Shuryak:2006ap} with the experiments on BEC of atoms in
traps.
The further development of the approach may provide qualiative insights
into the phases of QCD at high densities like the recently suggested
quarkyonic phase
\cite{McLerran:2007qj,McLerran:2008ua,Abuki:2008nm}.
Possible evidence for a triple point related to this new phase
comes from hadron production in heavy-ion collision experiments
\cite{Andronic:2009gj} to be further investigated at upcoming dedicated
facilities, e.g., CBM @ FAIR Darmstadt, NICA @ JINR Dubna.

\section{Formalism}
\subsection{Model Langrangian and mean field approximation}
Our starting point is a NJL-type Lagrangian for three colors and two flavors,
motivated from Fierz transformed one-gluon exchange
\begin{multline}
\mathcal{L} = \bar{q}(  {\rm i}\partial_\mu\gamma^\mu - m_0)q
+ G_S\big[\left(\bar{q}q\right)^2
+\left(\bar{q}{\rm i}\gamma_5{\bf \tau}q\right)^2\big]
\\
+ G_D \sum_{A=2,5,7}(\bar{q} {\rm i}\gamma_5 C \tau_2 \lambda_A \bar{q}^T )
       (q^T {\rm i}C\gamma_5 \tau_2 \lambda_A  q)~,
\end{multline}
with $C={\rm i}\gamma_2\gamma_0$ being the charge conjugation matrix,
$\tau=(\tau_1,\tau_2,\tau_3)$ and $\tau_i$ the Pauli matrices in flavor space
and $\lambda_A$ the anti-symmetric Gell-Mann matrices in color space.
The parameter choice $m_0=5$ MeV, $G_S\Lambda^2=2.1$ and $\Lambda=653$ MeV
reproduces the vacuum pion mass and decay constant.

For this model Lagrangian we can give the partition function in its bosonized
form
\begin{eqnarray}
	\mathcal{Z}=
\int {\mathcal D}\Delta^*{\mathcal D}\Delta {\mathcal D}\sigma {\mathcal D}\pi
\exp\left\{
-\int^\beta d^4x \frac{\sigma^2+\pi^2}{4G_S}+\frac{|\Delta|^2}{4G_D}
\right\}
\det S^{-1}~,
\end{eqnarray}
where diquark ($\Delta^*,~\Delta$) and meson ($\sigma,~\pi$) degrees of freedom
appear as collective fields instead of the quark ones which have been
integrated out leading to the determinant of the quark propagator $S$ in
Nambu-Gorkov representation.
For details of the further calculation we refer to \cite{Blaschke:2008uf}.
By minimizing the thermodynamical potential, we obtain gap equations which need
to be solved self-consistently.
The corresponding order parameters then characterize the phase structure of
the model.

\subsection{Gaussian fluctuations and polarization matrix}
We expand around the mean field Nambu-Gorkov quark propagator  
$S_{MF}
	\equiv
	\Bigl(
	\begin{smallmatrix}
		G^+&
		F^-
		\\
		F^+&
		G^-&
	\end{smallmatrix}
	\Bigr)~,
$ up to second order in the  matrix
$
	\Sigma
	\equiv
	\Bigl(
	\begin{smallmatrix}
		-(\sigma+\pi^+)&
		\delta^-
		\\
		\delta^+&
		-(\sigma+\pi^-)&
	\end{smallmatrix}
	\Bigr)~,
$
and obtain 
\begin{eqnarray}
	\ln\det S^{-1}
	=
	\Tr\ln S_{MF}^{-1}
	+
	\Tr\Big(S_{MF}\Sigma-\frac{1}{2}S_{MF}\Sigma S_{MF}\Sigma\Big)
	+ {\mathcal O} (\Sigma^3)~.
\end{eqnarray}
The fluctuations of the collective fields can be decomposed according to: 
$\pi^{+/-}\equiv {\rm i}\gamma_5\tau^{/t}\cdot\pi$,
$\delta^{+/-}\equiv {\rm i}\gamma_5\tau_2\lambda_2\delta^{*/}$,
and their amplitudes can be arranged in the vector  
$\vec{\phi} \equiv \{\pi,\sigma,\delta, \delta^{*}\}$.
Performing the trace operations over Nambu-Gorkov, flavor, color, Dirac and
momentum space we introduce the elements of the polarization matrix
$\Pi(k_0,{\bf k})$
(for details see \cite{Zablocki:2008sj,Blaschke:2010})
\begin{eqnarray}
  \frac{1}{2}\Tr\Big(S_{MF}\Sigma S_{MF}\Sigma\Big)=\phi_i \Pi_{ij} \phi_j~,
\end{eqnarray}
with $i,j=\{\pi,\sigma,\delta,\delta^*\}$ denoting the channels.
Some matrix elements are pairwise equal, e.g.,
$\Pi_{\sigma\delta}=\Pi_{\delta^*\sigma}, 
\Pi_{\delta\sigma}=\Pi_{\sigma\delta^*}$
and for real $\Delta$ even $\Pi_{\delta\delta}=\Pi_{\delta^*\delta^*}$.
Thus, the pions are degenerate, as expected for isospin symmetric matter.
We explicitly include the mixing terms between the $\sigma$ and the diquarks 
in our investigation, which has been omitted in the literature so far
\cite{Blaschke:2004cs,Ebert:1992ag,Ebert:2004dr}.
Performing the gaussian path integral over the fluctuation fields results in 
an expression for the thermodynamical potential
\begin{eqnarray}
\Omega(T,\mu)=-T\ln Z
=\ln\det S_{MF}
+\ln \det\left[\delta_{ij}/(2G_i)-\Pi_{ij}(\omega,{\bf k})\right]~,
\label{BSE}
\end{eqnarray}
where  $G_i=\{G_S,G_S,G_D,G_D\}$.
The mass spectrum of quasiparticle modes can be found from the condition
of the vanishing determinant in Eq. (\ref{BSE}) at ${\bf k}=0$ for
$\omega=\omega_i({\bf k}=0)=\{m_\pi,m_\sigma,m_\delta-\mu,m_{\delta^*}+\mu\}$.

\section{Results and Discussion}

\begin{figure}[htbp]
\includegraphics[width=.4\textwidth,angle=-90]{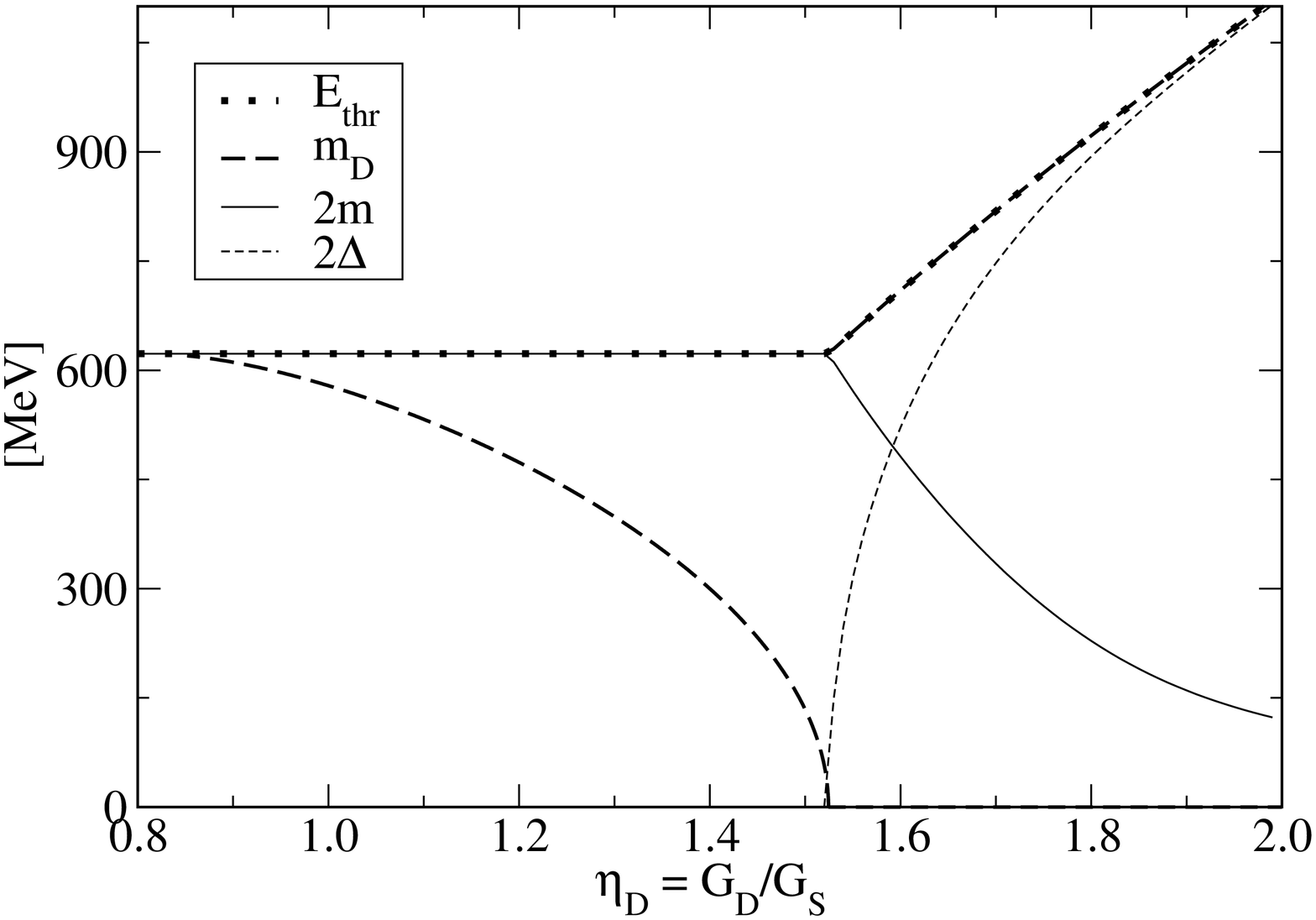}
\includegraphics[width=.4\textwidth,angle=-90]{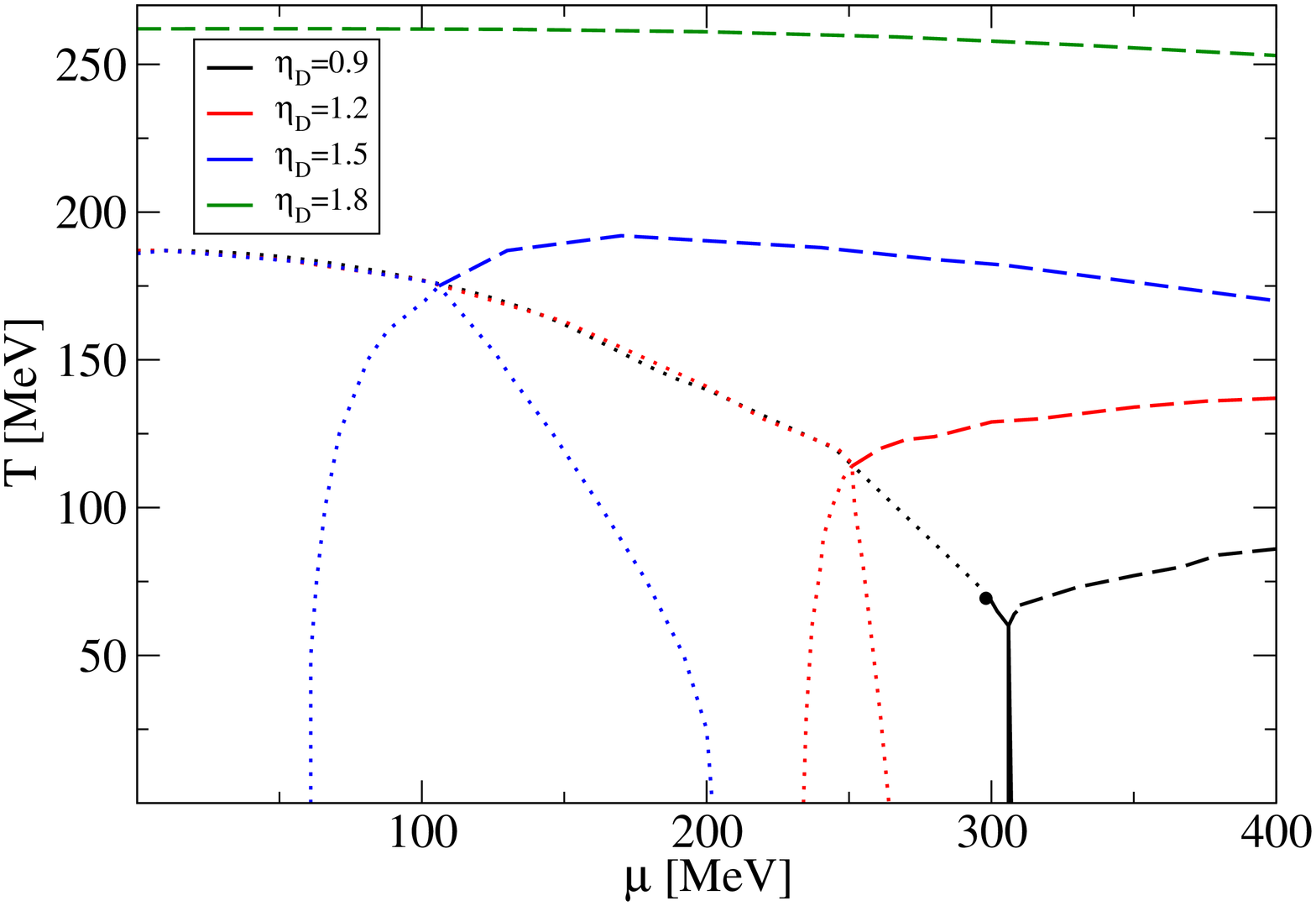}
\caption{
{\it Left panel:}
Vacuum diquark masses $m_D$ and threshold energies
$E_{\rm thr}=2\sqrt{m^2+\Delta^2}$ as function of the
coupling strength $\eta_D$.
{\it Right panel:}
Phase diagram for different values of the coupling strength
$\eta_D=0.9,1.2,1.5,1.8$.
First and second order phase transitions are indicated by solid and dashed lines respectively.
Dotted lines denote crossover transitions.
The black dot for $\eta_D=0.9$ indicates the critical endpoint for first order phase transitions.
For the discussion see text.
}
\label{figures}
\end{figure}

We want to discuss first the vacuum case $\mu=T=0$, where diquark and
anti-diquark are degenerate, the general discussion will be given in \cite{Blaschke:2010}.
In the normal phase, for the dimensionless diquark coupling strength 
$\eta_D=G_D/G_S$ in the range \cite{Sun:2007fc}
\begin{eqnarray}
\frac{\pi^2}{4G_S\big(\Lambda\sqrt{\Lambda^2+m^2}
+m^2\ln \frac{\Lambda+\sqrt{\Lambda^2+m^2}}{m}\big)}
\le\eta_D\le
\frac{3}{2}\frac{m}{m-m_0}=\eta_D^*~,
\end{eqnarray}
the polarization matrix is diagonal and thus \eqref{BSE} for the
diquarks gives
\begin{eqnarray}
1=8G_D\int\frac{{\rm d}^3p}{(2\pi)^3}\bigg(\frac{1}{E_{\bf p}+m_D/2}+\frac{1}{E_{\bf p}-m_D/2}\bigg)~.
\end{eqnarray}
If $\eta_D>\eta_D^*$ the matrix is not diagonal anymore.
Neglecting the mixing terms, we get two solutions for \eqref{BSE}, namely
\begin{eqnarray}
1&=&8G_D\int\frac{{\rm d}^3p}{(2\pi)^3}\bigg(
\frac{1}{E_{\bf p}^\Delta+m_D/2}+\frac{1}{E_{\bf p}^\Delta-m_D/2}
\bigg)
\\
1&=&4G_D\int\frac{{\rm d}^3p}{(2\pi)^3}
\bigg(1+\frac{E_{\bf p}^2-\Delta^2}{E_{\bf p}^2+\Delta^2}\bigg)
\bigg(\frac{1}{E_{\bf p}^\Delta+m_D/2}+\frac{1}{E_{\bf p}^\Delta-m_D/2}\bigg)
\end{eqnarray}
where $E_{\bf p}^\Delta=\sqrt{E_{\bf p}^2+\Delta^2}$, 
$E_{\bf p}=\sqrt{{\bf p}^2+m^2}$.
A Goldstone mode $m_D=0$ solves the first equation, which in this case
coincides with the gap equation for the pairing gap.
The solution of the second equation gives a massive mode.
The results are shown in the left panel of Fig.~\ref{figures} as a function
of $\eta_D$.
In the right panel of Fig.~\ref{figures} the phase structure of the
model is shown for four cases of coupling strenths: $\eta_D=0.9,1.2,1.5,1.8$.
While for $\eta_D<0.9$ there is no coexistence of chiral symmetry breaking and
diquark condensation, in the range $0.9<\eta_D<\eta_D^*$ one obtains
Bose condensation of bound diquarks in such regions of coexistence.
At $\eta_D>\eta_D^*$ a still more spectacular effect occurs: the vacuum state
itself is a Bose condensate of diquarks!
While this model description of a relativistic Fermi system at arbitrary
coupling is surely of methodological interest in the context of experiments
with Bose condensates of atoms in traps, its relevance for the discussion of
the phase structure of QCD requires a careful analysis of the
corresponding hadron spectrum.

{\bf Acknowledgements: }
D.Z. acknowledges the Heisenberg-Landau program for supporting his stay
at the JINR Dubna where most of this work has been carried out.
D.B. and D.Z. received support from the Polish Ministry of Science and Higher
Education (MNiSW) under grant No. NN 202 231837.
D.B. received funding from the Russian Fund for Basic Research under grant
No. 08-02-01003-a. This work was supported in part by CompStar, a Research
Networking Programme of the European Science Foundation.

%
%

\end{document}